\documentclass[12pt, draftclsnofoot, onecolumn]{IEEEtran}

\usepackage{amsmath}
\usepackage{graphicx}
\usepackage{subfigure}
\usepackage{multicol} 
\usepackage{epstopdf} 
\usepackage{epsfig} 
\usepackage{multirow}
\usepackage{bm}
\usepackage{array}
\usepackage{color}
\usepackage{float}
\usepackage{amsfonts}
\usepackage{caption}
\usepackage{chemformula}

\usepackage{afterpage}

\begin{document}



\title{Deep Learning for Visual Neuroprosthesis}

\author{\IEEEauthorblockN{{Peter Beech}\ddag, Shanshan Jia\dag, Zhaofei~Yu\dag, Jian~K.~Liu\ddag \\}
	\small{\IEEEauthorblockA{
    \ddag University of Leeds 
    \dag  Peking University
        }
        }
}

\maketitle

\begin{abstract}

The visual pathway involves complex networks of cells and regions which contribute to the encoding and processing of visual information. While some aspects of visual perception are understood, there are still many unanswered questions regarding the exact mechanisms of visual encoding and the organization of visual information along the pathway. This chapter discusses the importance of visual perception and the challenges associated with understanding how visual information is encoded and represented in the brain. Furthermore, this chapter introduces the concept of neuroprostheses: devices designed to enhance or replace bodily functions, and highlights the importance of constructing computational models of the visual pathway in the implementation of such devices. A number of such models, employing the use of deep learning models, are outlined, and their value to understanding visual coding and natural vision is discussed.

\end{abstract}

\section{Introduction }

The ability to perceive and make sense of the surrounding environment is an essential task for all organisms. Humans achieve this through a multitude of senses involving chemical and mechanical receptors, yet perhaps the most important source of information about the environment comes from visual information. The human eye detects patterns of photons deflected by the physical form of the surrounding environment in order to perceive their existence, yet the fundamental processes by which this photon pattern information is encoded and represented within the biological brain as the basis of visual perception remain a mystery despite decades of research. 

A neuroprosthesis is a device designed to substitute for a part of the human body in order to restore motor, sensory, or cognitive ability. This is commonly achieved through direct interaction with neuronal activity and electrical action potentials within the brain - the primary coordinator of such processes. Sensory information is normally encoded into electrical action potentials, the language recognizable by the human nervous system, via computations executed by networks of cells. Thus, a thorough understanding of this encoding process is critical to translate non-biological information into a form recognizable to the biological body, and vice versa, allowing integration of non-biological devices with the biological human body to reproduce healthy sensory-mediated body functions \cite{Yu2020, Nirenberg2012}. Both those devices sending information, such as retinal neuroprosthetics, and those receiving commands, including neuroprosthetics modulating movement, must consider this encoding process in order to integrate successfully with the nervous system. 

Recent advancements in the fields of motor and cochlear neuroprostheses have been comparatively successful in restoring near-healthy sensory ability to patients, in contrast to current retinal neuroprosthetics, which are only partially able to restore natural vision \cite{Nirenberg2012, Gilja2012, Seeber2016}. The difference between restorative ability of different sensory neuroprostheses may be significantly attributed to the number of receptive cells, and the complexity of encoding performed within the sensory organs. Although encoding of the visual signal was previously believed to occur exclusively within the brain, the retina is now known to encode a large degree of vision, exhibiting a multi-layered architecture of tightly coupled and highly interactive cells \cite{liu_inference_2017, Liu2022, Gollisch2010, Euler2014, liu_spike-triggered_2015}, unlike the more simple motor and cochlear organ architectures. Hence, the retina, with a far higher density and number of receptors than those involved in motor or audial response, proves a far more difficult challenge to artificially reproduce. As such, in addition to the complex encoding of sensory information within the brain which is performed for all types of sensory information, the utilization of retinal neuroprostheses requires additional encoding steps within the visual organ to be considered. 

Some stages of visual perception are understood to an extent, yet there remains mystery around when, where, and how different aspects of visual information are encoded within the pathway for visual perception. Common hypotheses suggest encoding and decoding take place following two major concepts: hierarchical processing, and functional specialization \cite{GrillSpector2004}. Hierarchical processing is the idea that visual perception is achieved through simple initial representation of information, and gradual transformation into more abstract, holistic, and multimodal representations \cite{DeYoe1988}. Function specialization proposes several specialized pathways exist, each processing different aspects of the visual field. The pathway is thought to encode a variety of visual field features including orientation and contours, motion and direction, stereoscopic depth, and color-based, texture-based, and shape-based features.

\section{Visual Pathway}

The human visual pathway begins at the eye, where light information is detected. It is then processed along hierarchical pathways involving the retina, subcortical regions, then cortical areas of the brain. Processed visual information is distributed to further regions of the brain where it may be applied, yet there is little to no evidence suggesting how humans are able to consciously perceive vision. Visual information processing is thought to involve multiple pathways, of which the most well established are the ventral and dorsal streams. The ventral stream, termed the ``what'' stream, is thought to be involved in perception, and object and form recognition \cite{Mishkin1983, GrillSpector2004}. Whereas, the dorsal stream, also referred to as the ``action" or ``where" stream, is associated with visually guiding behavior, and visuospatial judgment \cite{Goodale1991, Mishkin1983, GrillSpector2004}.

    \subsection{Retina}
    
Within the eye, the retina is responsible for the translation of visual information in the form of photons into a code of electrical spikes within neuronal cells that are recognizable to the biological body. Visual information was previously believed to be sent to the brain as a pixel-based representation, however, it is now clear that a significant amount of pre-processing of the visual signal occurs in the retina before it is sent to subcortical regions of the brain \cite{Kim2021}. The layered structure of the network of cells within the retina can be roughly divided into three layers, consisting of 5 main categories of neuronal cells \cite{Yu2020}. Visual information flows from photoreceptors, to bipolar cells, and finally to ganglion cells, with inhibitory horizontal and amacrine cells thought to modulate signals between and throughout layers.

    \subsection{Photoreceptors}
    
Two light-sensitive cells in the retina: rods and cones, are responsible for translating light signals into biological electrical activity. Photons induce temporary chemical changes to proteins called rhodopsins and opsins, within rod and cone photoreceptors respectively, causing a graded change in electrical potential of the cell membrane. This change in electrical potential mediates the release of neurotransmitters, chemical signaling molecules, into the synapse and onto postsynaptic cells deeper below the surface of the retina. A large amount of the processing within the retina is thought to be mediated by graded potentials rather than action potentials \cite{Purves2001}. Hence, contrary to most other types of sensory receptor cells, increasing light intensity causes membrane hyperpolarisation rather than depolarisation and a resulting action potential. The resting potential of the membrane is roughly -40mV under dark lighting conditions, becoming progressively more negative under increasing illumination up to a hyperpolarisation of -65mV.  Nevertheless, similarly to other types of nerve cells, presynaptic neurotransmitter release is mediated by, and positively proportional to, intracellular calcium ion concentration. More voltage-gated calcium ion channels are open in the dark, causing greater calcium ion concentration, and neurotransmitter release when not illuminated. Hence, in addition to the absolute signal intensity, some information conveyed by photoreceptors may be held in changes in the signaling strength.

    \subsection{Horizontal cells}
    
Horizontal cells hyperpolarise in response to light, and are thought to be responsible for lateral inhibitory interactions which modulate photoreceptor activity as it is transmitted to bipolar cells \cite{Kaneko1970}. These lateral interactions are proposed to contribute to the center-surround receptive field structure of retinal ganglion cells, responsible for the ability to detect contrasts in luminance over a wide range of light intensities \cite{Purves2001}. The primate retina is thought to include two types of horizontal cells, yet their functional contribution to the retinal output is not fully understood \cite{Chaya2017, Kim2021}.

    \subsection{Bipolar cells}
    
Photoreceptor cells interface directly, or indirectly via horizontal cells, with bipolar cells, which in turn synapse with other adjacent horizontal or bipolar cells \cite{Purves2001}. As their name suggests, bipolar cells can be functionally classified according to receptive field component response polarity into two types: on and off type center-surround, responding inversely to spot-shaped illumination \cite{Kaneko1983}. Around 12 types of bipolar cell subtypes have been determined within the primate retina, speculated to have different functional roles in visual coding \cite{Tsukamoto2017, Kim2021}. Bipolar cells were previously believed to exhibit graded signaling alike the previous retinal cells. However, bipolar cells have recently been observed in fish to be responsible for conversion of a graded analog stimulus into a spike-based electrical form, thus suggesting this as one of the possible functions of the bipolar cell \cite{Baden2011}.

    \subsection{Amacrine cells}
    
Signals from bipolar cells are received and modulated by amacrine cells \cite{Purves2001}. Amacrine cells, similarly to horizontal cells, exhibit inhibitory lateral interactions. They have been hypothesized to exhibit multiple functions, with more than 25 types of amacrine cells categorized in the primate retina \cite{Kim2021}. Amacrine cells are thought to generate object motion sensitivity and textured edge detection from nonlinear representations of speed and texture respectively, originating from bipolar cell input \cite{Jadzinsky2013}. However, the functions of more than 20 types of amacrine cells are yet unknown.

    \subsection{Ganglion cells}
    
Ganglion cells receive signals from amacrine cells. In 1953, Kuffler demonstrated the response of concentric on-off surround receptive fields in retinal ganglion cells to spots of light \cite{Kuffler1953}. Previously treated as a relatively uniform population consisting of on or off center surround types, retinal ganglion cells have been shown to display a diverse range of subtypes, with unique morphological features, functions, and related pathways \cite{Kim2021}. The retinal ganglion cell population, thought to consist of more than 18 different types of ganglion cell, convey in parallel an array of attributes of the visual field. These attributes are derived from the unique pathways associated with each ganglion cell type that spans from the photoreceptors to the ganglion cells. Known attributes include spatial contrast, color, motion and direction, flicker, fine and coarse textures, and absolute light intensity, but which and how features are encoded is a highly debated topic; thus, other undiscovered features are likely also encoded.

    \subsection{Subcortical regions}
    
Visual information from each retina is transported via the optic nerve, consisting of temporal and nasal fibers corresponding to respective sections of the visual field \cite{Gupta2022}. These fibers intersect at the optic chiasm, located directly behind and above the pituitary stalk, where over half of the nasal fibers from each eye cross over to join the temporal fibers from the opposite eye. Axons of the two resulting optic tracts, carrying visual information about the left and right sides of the overall visual field, synapse mostly onto the lateral geniculate nucleus (LGN); a few axons thought to control parasympathetic pupillary innervation synapse onto the superior colliculus and Edinger-Westphal nuclei.

    \subsection{LGN}
    
The LGN is an extension of the thalamus consisting of 6 layers, 4 parvocellular and 2 magnocellular, receiving information from parasol and midget ganglion cells respectively. Each fiber innervates alternate layers of the LGN, totaling 2 parvocellular and 1 magnocellular per fiber \cite{Gupta2022}. Each layer of the LGN preserves a retinotopic map of inputs to the retina, with each of the 3 layers mapping the entire visual hemifield. Damage to the parvocellular layers of the LGN in monkeys significantly impairs visual acuity and color perception, while having little effect on motion perception, suggesting the importance of parvocellular layers in the ventral stream. Whereas, damage to magnocellular layers has no effect on visual acuity and color perception, but impairs motion perception, suggesting a separate role related to the dorsal stream of visual perception. Around 90 percent of the inputs to the LGN are modulatory inputs from the cortex and brainstem, with only about 10 percent originating from the retina, thus it has been suggested that the LGN may not do much processing of the visual information, but rather, is involved in both combining inputs from both visual hemifields, and the modulation of inputs, so that such processes do not need to occur within the very compact structure of the retina.

    \subsection{Visual cortex}
    
The human visual cortex is located in the occipital lobe of the cortex, and divided into 5 regions termed V1, V2, V3, V4, V5, which process visual information from the LGN along the visual pathway \cite{Huff2022}. Visual information from the LGN is initially primarily directed to the primary visual cortex (V1), also known as the striate cortex due to its anatomically distinctive bands of myelinated axons under microscope, and is the most well understood aspect of the visual cortex. Divided into 6 distinct layers with multiple cell types and functions, V1 primarily responds to simple visual components, such as orientation and direction; Thus, virtually all visual processing tasks activate V1. Simple cells, found primarily in V1 of the visual cortex, respond to specific visual features, including the orientation of edges and lines. The fourth layer of V1 receives information from the LGN, and correspondingly has been shown to express the highest concentration of simple cells. Layers two, three, and six exhibit more complex cells, which do not appear to represent single receptive fields, and instead respond to the summation of multiple simple cell receptive fields. In 1959, Hubel and Wiesel famously demonstrated orientation and direction selectivity of single neurons in the cat V1 to visual oriented bar stimuli, and later in monkeys \cite{Hubel1959, Hubel1968}. They discovered the presence of on and off areas in V1 as in retinal ganglion cells, but noted differences in arrangement and shape of on and off areas from the concentric arrangement in the retina, indicating differences in the organisation of visual information as it progressed along the visual pathway. 

A large amount of evidence indicates the visual pathway known as the ventral stream to mediate visual recognition in the primate cortex, involving V1, V2, V4, and further occipito-temporal regions \cite{DiCarlo2012}. Hence, it can be hypothesized that areas along this hierarchical pathway retain similar visual information, but it may be arranged differently in order for different regions to extract different aspects of visual feature information. Cells in V2 receive integrated information from V1, and have been shown to respond to illusory contours in images \cite{Peterhans1991, Huff2022}. V2 has been shown to exhibit feedback connections with V1, and feedforward connections with V3, V4, and V5 \cite{Huff2022}. The extrastriate cortex includes areas V3, V4, and V5, and is proposed to be involved in more abstract, higher levels of analysis due to response of cells within more specific sets of visual features than V1 \cite{VanEssen1979}. Cells in V4 respond only to specific patterns or colors of stimulus. Evidence suggests V4 to have characterized selectivity to contours, using simple parameterized feature spaces, and uses local image features to perceive texture in natural images; yet, it is unknown how much of this information may be inherited from processing in V2 \cite{Ziemba2015, Okazawa2015}. The inferotemporal cortex, and other regions appearing near the top of the ventral stream hierarchy appear to respond best to complex stimuli such as natural scenes, and have been indicated to hold accurate and robust information about object identity and category, in addition to some information about object position and scale \cite{Hung2005}. In occipito-temporal regions, the cells have been shown to respond selectively only to particular shapes \cite{Desimone1984, Logothetis1995}.

\section{Neuroimaging methods}

Recent advancements in neuroimaging techniques, particularly noninvasive functional imaging methods, have contributed to our understanding of the functional organization of the human visual cortex, allowing mapping and analysis of human cortical visual areas \cite{GrillSpector2004}. This knowledge opens the possibility for use of animal models with functionally homologous regions to humans, allowing investigation of regions with more invasive methods. The measurement of neural signals is a requirement for the study of coding within the brain, yet this feat is extremely difficult, thus hard to achieve without making significant compromises. Neuroimaging is the field of investigating how these signals can be recorded or predicted, and describes a vast array of methods that may be used to observe brain structure and function. 

Neural activity in the human brain is accompanied by several other biological changes which may be measured more easily than the intracellular electrical potentials of neurons. As such, each neuroimaging method has advantages and corresponding disadvantages depending upon what is being observed. Current experiments often use a combination of multiple methods to cover the weaknesses of each method. However, while most methods allow imaging of the outermost parts of the brain, deeper and subcortical regions of the brain are more difficult to image. Multimodal imaging of regions such as the thalamus, a part of the human visual pathway, is therefore more difficult to achieve. Methods may either be non-invasive or invasive, with the latter requiring surgical procedures to directly access brain tissues. As such, invasive imaging is rarely performed on human participants; instead, animals thought to share neural resemblance, such as primates or other mammals, are used as models of the human brain. For example, rats and mice are mammals which, similarly to humans, rely primarily on vision. They have large eyes, high visual acuity, similar parallel visual pathways in the thalamus, and ocular dominance and orientation columns in the visual cortex \cite{Kaas2001, Nassi2009}. Primates, which share a relatively recent common ancestor with humans, generally differ in higher order cortical visual regions, but remain similar at lower hierarchical tiers including areas V1, V2, and V5 of the visual cortex \cite{Tootell2003, Kaas2001}. 

Commonly, a simple artificial stimulus, such as an oriented bar, is presented to a subject \cite{Hubel1959}. Subsequent observation of neuron spike responses suggests a role of the observed population in the pathway encoding representative features of the stimuli; in the case of an oriented bar, features include edge detection and orientation discrimination. Individual cells of populations involved in the visual pathway often display specific tuning to the attributes of visual features. For example, the primate primary visual cortex displays a retinotopic feature map exhibiting a pinwheel-like pattern of orientation preference \cite{Koulakov2001}. Nevertheless, the definite functional benefit to the natural occurrence of such patterns is unknown. Such retinotopic feature maps are also present in other regions in the hierarchical pathway in some animals, yet others, such as rodents, have been shown to display orientation preference without a specific spatial map of organization \cite{VanHooser2005}. Nevertheless, this orientation preference is thought to be preserved within several regions within the visual hierarchy, with recent observation of around ~25 percent directional modulation in the mouse hippocampus with respect to visual queues, despite previous research indicating no directional selectivity in mouse hippocampus \cite{Acharya2016}. 

The representation of information within spike trains is still under debate, with several possible methods for information to be carried within trains \cite{Brette2015}. The most obvious form is in the firing rate of spikes, however information could also be held in the timing between spikes. Furthermore, the correlation of spike timing between multiple neurons in a population may similarly convey information. Additionally, it is not known exactly how much redundancy is built into the visual system, though initial estimates predict redundant information is removed as information moves along hierarchical pathways \cite{Petras2021}.

    \subsection{fMRI}
    
In the study of visual coding, a wide range of methods are employed, from directly observing electrical changes, to the use of other biological changes occurring within the brain which may be used as indicators of activity. For example, in order to meet the high metabolic demands of neurons in the brain without the space for storage of energetic resources, an adequate supply of these resources must be maintained by an increase in cerebral blood flow (CBF) proportional to the increase in local energy demands - in a phenomenon termed neurovascular coupling. This particular relationship is the foundation of a non-invasive functional brain imaging method used to map neuronal activity in the brain by measuring changes caused blood oxygenation to magnetic field distortions; this method is known as functional magnetic resonance imaging (fMRI) \cite{DeYoe1994}. However, as fMRI does not measure the actual electrical activity of neurons, the blood oxygen dependent (BOLD) signal recorded is not a measure of absolute neuronal activity, and instead requires comparison between regions in order to extract value from measurements. Thus, a region known not to be involved with the pathway of investigation, from previous neuroimaging investigation, must also be imaged alongside the region of interest in order to provide a baseline for comparison, controlling for changes such as overall blood volume; this makes fMRI less useful as a tool for discovering new regions in pathways, and more of a tool for investigating known pathways. Additionally, haemodynamic changes occur within a few seconds after neural activity, and thus the recorded signal is not a temporally synchronized representation of neural activity. Furthermore, fMRI is expensive in both equipment and computational resource requirements. Layers of the brain are imaged one at a time, thus the operator must balance between imaging refresh rate, acuity and area. High resolution fMRI can only be performed in smaller regions, with lower resolutions and imaging areas required for shorter imaging time steps. Nevertheless, fMRI offers a unique possibility to non-invasively observe activity in deeper regions of the brain with reasonably high spatial accuracy.
A previous study fairly successfully employed the use of fMRI imaging of the visual cortex to predict novel natural movie scenes by comparing the BOLD signal to that measured in previously presented movies \cite{Nishimoto2011}. Hence, demonstrating the BOLD signal, dependent on spiking, over the whole human visual cortex is representative of the overall visual information presented at the retina. Yet, it is difficult to predict which regions, or voxels of the fMRI recordings, provide which types of visual information. Moreover, this method of reconstruction requires previous recordings from the individual to tune the decoding model, rather than being built on an understanding of the visual code represented by spikes in the visual cortex.

    \subsection{Two-photon calcium imaging}
    
A fluorescent marking technique termed two-photon calcium imaging observes changes to the intracellular calcium concentration as fluorescence, mediated by genetically modified proteins \cite{Stosiek2003, Vries2020}. However, in addition to gene modification, invasive procedure is also a requirement to visually observe the brain tissue, thus use is limited to non-humans, and it is only able to image layers near the brain surface. Nevertheless, the ability of two-photon calcium imaging to simultaneously observe entire cell populations with high single cell spatial accuracy is useful for investigation of visual coding at a cellular level. In order to transmit information between neuronal cells at synapses, neurons release chemicals called neurotransmitters. The release of neurotransmitters is mediated by the voltage-mediated influx, and subsequent increase in concentration, of calcium ions within the presynaptic axon. Less negative membrane potentials increase the flow of calcium ions into the cell, temporarily increasing intercellular calcium ion concentration. Calcium imaging observes the changes in this concentration via introducing genetically modified fluorescent proteins to the neurons; when calcium ions bind to these proteins, they fluoresce, thus causing cells with increased calcium ion concentration, corresponding with increased neurotransmitter release into the synapse, to fluoresce. Nevertheless, genetic modification to cause expression of such proteins is limited to non-humans, is time-consuming, and may cause other aspects of biology to diverge from the typical behavior. Thus, care must be taken to avoid genetic lines which may affect the object of research. Furthermore, similarly to the change in blood oxygenation, the change in calcium ion concentration is slower than directly reading electrical action potentials, thus imaging incurs some temporal lag.

    \subsection{Electrophysiology}
    
Electrophysiology includes methods that measure electrical changes within cells, or in the surrounding tissues. Electroencephalography (EEG) is a common procedure applicable in humans for the diagnosis of epilepsy, and measures the voltage fluctuations caused by electrical activity \cite{Biasiucci2019}. Extracranial EEG is a non-invasive method whereby the outer surface of the head is covered with electrodes, however due to the physical barriers of the skin and skull between the electrodes and brain tissues, electrical signals are far smaller and less spatially accurate than other imaging methods, making individual cell response measurements unattainable. Intracranial EEG involves invasive surgical procedure to insert electrodes within the brain in vivo, thus is performed in patients primarily for clinical, rather than scientific, interest, providing greater signal strength and accuracy than extracranial EEG. Similarly, electrophysiological methods used in non-humans involving multi-electrode arrays require direct access to brain tissues, but provide good spatial and temporal accuracy. Large cell population electrode recordings for visual information decoding using intracellular electrophysiology are practically unfeasible in vivo due to the accuracy in which electrodes must be placed, but extracellular electrophysiology may be used to record cell populations simultaneously using multi-electrode arrays, or intracranial EEG \cite{Biasiucci2019, Siegle2021}.

\section{Deep learning models}

\begin{figure}[tbp]
\centering 
\includegraphics[width=1\textwidth]{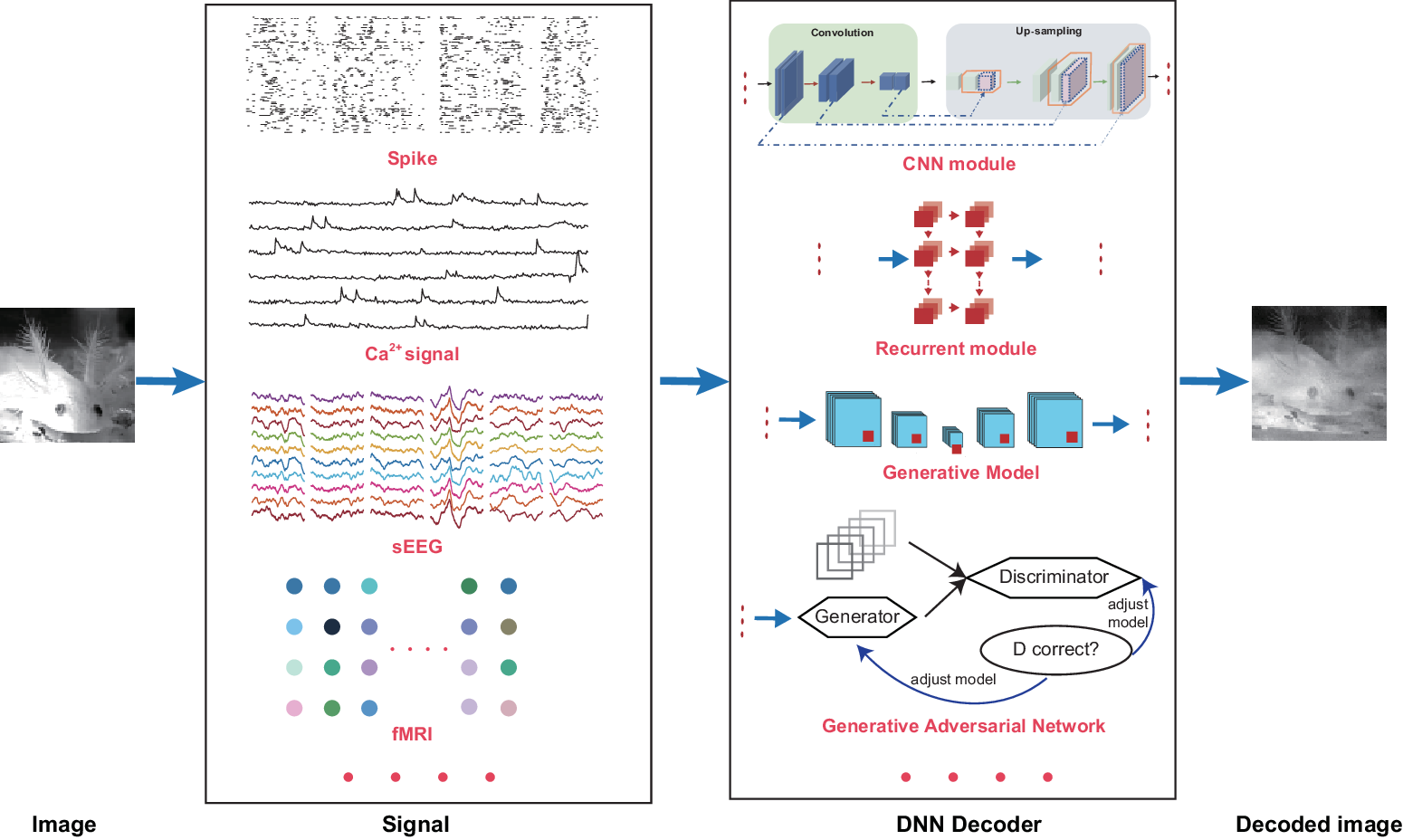} 
  \caption
  {Overview of the deep learning framework for decoding visual scenes from neural signals of visual systems.
  }
\label{Fig.1} 
\end{figure}

In recent years, deep neural network (DNN) has become a powerful model in computer vision tasks. By training a large set of natural images for deep learning, its visual object recognition can achieve human-like performance~\cite{lecun_deep_2015}. Therefore, using deep learning methods to explore visual features of visual systems has become an influential and popular research direction. The research on deep learning-based decoding of neural signals mainly includes studies on convolutional neural networks (CNN), recurrent neural networks (RNN), generative models, generative adversarial network (GAN), and semi-supervised models(~Figure~\ref{Fig.1}). These models are starting to play an important role in dealing with natural visual scenes for improving the functionality of visual neuroprothesis~\cite{Yu2020, shah_computational_2020, Gogliettino2023, Zhang2022a}.

\subsection{Convolutional Neural networks}
Convolutional neural networks driven by image recognition have been shown to be able to explain the retinal response of natural images and videos~\cite{Yan2022, zheng_unraveling_2021}. In relevant research on decoding, it has been found that CNN can reliably predict and decode visual stimuli (natural images or even natural videos) by learning the input neural signals, although it lacks any mechanism to explain temporal dynamics or feedback processing ~\cite{horikawa2017generic,seeliger2018convolutional,wen2018neural}. This type of model is mainly used to decode input stimulus categories based on neural responses.

The decoding model for decoding stimulus categories normally consists of a CNN module and a fully connected module. There are generally many options for implementing CNN models, for example, the model framework can be a 5-layer convolutional network~\cite{horikawa2017generic}. In addition, one can also use AlexNet, VGG, ResNet and other more complicated and multilayered networks to implement the CNN module, which mainly uses big data sets (such as ImageNet) for pre-training~\cite {seeliger2018convolutional, wen2018neural, qiao2019category}. The encoding and decoding model describes the relationship between CNN and the brain in terms of fMRI signals~\cite{wen2018neural}, where a pre-trained AlexNet was employed to extract hierarchical visual features from natural stimuli. Here we can use this model as an example explaining how it works for decoding. This typical example used a CNN consisting of 8 stacked component layer architecture computing units: the first 5 layers are regular layers and the last 3 layers are fully connected for image object classification. The image input is fed into the first layer; The output of one layer is used as the input of the next layer. Each convolutional layer contains a large number of units and a set of filters (or kernels), which extract filtered outputs at all positions from their inputs through a rectified linear function. The first to fifth layers are composed of 96, 256, 384, 384, and 256 kernels, respectively. Maximum pooling is implemented between layers 1 and 2, between layers 2 and 3, and between layers 5 and 6. For classification, layers 6 and 7 are fully connected networks. The 8th layer outputs a probability vector using a softmax function, which is used to classify the input image into various categories. The number of units from the 6th to 8th layers is 4096, 4096, and 1000, respectively. Each voxel response is treated as a specific pixel mode that drives the response. After decoding the model, the fMRI signal is directly decoded to estimate feature representations in the visual and semantic spaces, which are used for direct visual reconstruction and semantic classification, respectively. These results confirm, summarize, and extend previous findings and highlight the value of using deep learning as an ensemble model of the visual cortex to understand and decode natural vision.

In addition, some works are based on CNN models for stimulus reconstruction, particularly natural images and videos. Fully-connected network and autoencoder and Unet deep network framework to reconstruct visual stimuli based on two-photon responses recorded from macaque V1~\cite{zhang2022decoding}. Such a network framework is similar to the encoder-decoder module, which will be explained in detail in the Generative model later.

\subsection{Recurrent Neural Network}

The visual system is not a pure forward network, but contains a highly rich variety of connectivity patterns, which is destined to add new network structures beyond CNNs to further approximate brain states, such as RNNs. The study of visual cortex models emphasizes the role of circular connections in the visual processing of the model itself, which helps to ``fill in" missing data~\cite{kar2019evidence,kietzmann2019recurrence}.

In the human visual system, the visual cortex is functionally divided into ventral and dorsal streams, with the ventral cortex being primarily responsible for object detection. Anatomical studies have shown that the connections between the ventral cortex are bidirectional~\cite{bar2003cortical}. Bidirectional (forward and backward) connections provide anatomical structures for the bidirectional information flow in the visual cortex. Forward and backward information flows play different roles in recognition tasks. Thus, the experimenter can assume that bidirectional information flow carries semantic knowledge from the high-level visual cortex. Therefore, maximizing bidirectional information flow in the visual cortex is important for category decoding~\cite{qiao2019category}, which gives a model based on bidirectional recurrent neural network (BRNN) to decode corresponding stimulus categories from fMRI data. The forward and backward directions in the BRNN module represent the bottom-up and top-down approaches, including the encoding and decoding parts. For the encoding part, they obtain the corresponding features for a given image stimulus based on a pre-trained ResNet-50 model, and use these features to fit each voxel to construct a voxel-coded model. Based on the fit performance, they can measure the importance of each voxel for all visual regions. They selected a fixed small number of voxels with high predictive correlation for each visual region (V1, V2, V3, V4, and LO) to prevent overfitting in subsequent decoding. For the decoding part, they constructed an RNN module and used selected voxels from each visual region as the five nodes of the sequence input to exploit the hierarchical visual representation and bidirectional information flow in the visual cortex. Finally, they combine the extracted features of the bidirectional RNN module into the input of the last fully connected softmax classifier layer to predict the category. This approach can more effectively exploit hierarchical information representation and bidirectional information flow in the human visual cortex. Experiments have shown that this approach can improve the accuracy of three-level category decoding. Comparative analysis verifies and reveals that, due to bidirectional information flow, in addition to hierarchical, distributed and complementary representations consistent with previous studies, related representations of categories are also included in the visual cortex. The RNN module can also choose to model stacked gated loop units with skip connections, such as GRU or LSTM~\cite{yang2020decoding, zheng_unraveling_2021}.

In addition, there has been a lot of work to introduce attention mechanisms into RNN-based decoding models, which can be implemented through fully connected layers trained jointly with RNNs~\cite{qiao2019category,yang2020decoding}. The attention mechanism emphasizes task-related responses and reduces redundant information in the neural signals, thereby improving the accuracy of the model. In addition, dropout ~\cite{srivastava2014dropout} is added to acyclic connections to prevent overfitting, while early stopping methods are used to achieve better generalization performance of the model.

\subsection{Generative Model}

There is also the assumption that the brain is mostly unsupervised. AI has explored a similar concept of unsupervised learning. For example, recent research has also focused on the application of deep generation techniques in visual neural decoding, such as autoencoder and variational autoencoder (VAE)~\cite{zhang2020reconstruction,han2019variational,du2018reconstructing,Zhou2023}. 

VAE is also a recently popular Generative model, which goes beyond the paired matching (stimulus-response) of a given real physiological data set and uses an encoder to describe the probability distribution of the potential state space. New stimuli not present in real physiological datasets can be created by sampling from the latent state space, similar to the regular patterns of real stimuli. \cite{han2019variational} used VAE decoders to convert potential variables into video frames and reconstruct video inputs for functional magnetic resonance imaging activities. VAE is trained using a five-layer encoder and a five-layer decoder to learn visual representations from various unlabeled images. Using the trained VAE, they predicted and decoded cortical activity observed through fMRI in three human subjects who were passively watching a natural video. The difference in coding performance between CNN and VAE is mainly attributed to their different learning objectives rather than their different model architectures or parameter numbers. Despite the low encoding accuracy, VAE provides a more convenient strategy for decoding fMRI activity to reconstruct video inputs. First, the fMRI activity is converted into latent variables of the VAE, which are then converted into reconstructed video frames by the VAE decoder. This strategy has advantages over other decoding methods, such as partial least squares regression, in order to reconstruct the spatial structure and color of visual inputs. These findings highlight VAE as an unsupervised model for learning visual representations, as well as its potential and limitations in interpreting cortical responses and reconstructing naturalistic and diverse visual experiences.

Recently, a universal decoding framework called Spike Image Decoder (SID) based on deep learning neural networks was developed to perform end-to-end decoding processes from neural spikes to visual scenes~\cite{zhang2020reconstruction}. In contrast to previous studies, the proposed SID is able to achieve state-of-the-art performance for reconstructing natural visual scenes, including static images and dynamic videos, from simultaneous spikes of RGC populations recorded from separated animal retinas. The workflow of SID is to first map the collected ganglion spike signals to the intermediate image at the pixel level using a three-layer spike image transfer (Dense neural network) to map each RGC spike. The signal is then fed into an image-to-image autoencoder to further optimize the reconstructed image. In this work, an end-to-end impulse image decoder is proposed to reconstruct stimulus images from neural spiking processes based on retinal ganglion cells. Obtaining state-of-the-art reconstruction performance in both static and dynamic natural scenes. Moreover, the excellent performance of arbitrary visual scene reconstruction on popular image datasets is demonstrated by using an additional encoding model. In addition, SID can also be used to decode fMRI data, demonstrating superior performance compared to other methods developed in the context of fMRI decoding. Finally, the generalization ability of SID is tested by decoding various types of dynamic videos with pre-trained SID to achieve real-time encoding and decoding of neural spikes in visual scenes. This model is potentially useful for visual neuroprosthesis to quantify the quality of neural signals after treatment. In particular, one can embed this model into the newly developed neuroprosthesis based on nanowire~\cite{tang_nanowire_2018}, where the nanodevice can replace damaged photoreceptors and directly transcode light signals to downstream neurons. In this way, one can overcome the difficulty of manipulation of electrical stimulation used in conventional retinal neuroprostheses~\cite{Lu2022}.

\subsection{Generative Adversarial Network}

Recovering a complete image is complicated by the noisy and high-dimensional nature of neural representations, which contain incomplete information about image details. Therefore, reconstructing complex images from brain activity requires strong prior knowledge. Since GAN has achieved great success in synthesizing high fidelity images, many studies suggest training GAN to learn the image Generative model based on brain activity measurement~\cite{st2018generative,seeliger2018generative,guccluturk2017reconstructing,vanrullen2019reconstructing}.

\cite{seeliger2018generative} trained the depth convolution Generative adversarial network DCGAN on the image dataset to learn the potential state space, and then used it to generate handwritten characters and natural Grayscale from fMRI signals. They explored a method to reconstruct visual stimuli from brain activity. Using a large database of natural images, they trained a deep convolutional generative adversarial network capable of generating grayscale photos, similar to the stimuli presented in two fMRI experiments.

The model architecture is that the generator network consists of a linear layer and four deconvolution layers, and each layer is followed by a batch normalization and rectification linear Activation function (ReLU). The linear layer takes z and maps it to the first deconvolution layer of the desired 512 feature channels. The generator is then mapped to 256, 128, 64, and 1 feature channels on the deconvolution layer. In each deconvolution layer, the kernel size is 4 * 4 and the step size are 2. The final step is to apply tanh to the output value, ensuring that the pixel output is between [-1,1]. The discriminator network consists of four convolutional layers, followed by batch normalization and exponential linear activation (ELU). Gaussian noise with standard deviation 0.15 is added to the input image before the image is fed into the discriminator. Except for the initial layer (which has a 3 * 3 convolutional kernel), all layers use a 4 * 4 size convolutional kernel and a 2-step size. These layers map from 1 to 32, followed by 64, 128, and 256 feature channels, and then a linear layer that maps all final activations to a single value reflecting the discriminator's decision.

The model learns to predict the latent space of the generative model from the measured brain activity. The goal is to create an image similar to the presented stimulus image using the previously trained generator. Using this approach, they were able to reconstruct the structure and some semantic features of a portion of the natural image set. Behavioral tests showed that in a Pairwise comparison of two natural image datasets, subjects were able to recognize the reconstruction of the original stimulus in 67.2\% and 66.4\% of cases, respectively. Their approach does not require end-to-end training of large Generative models on limited neuroimaging data. Rapid developments in generative modeling are expected to further improve the reconstruction performance. In addition, there are many GAN-based facial image decoding methods~\cite{guccluturk2017reconstructing,vanrullen2019reconstructing}. For example, GAN was used to train VAE on large celebrity face datasets and obtain corresponding Potential space~\cite{vanrullen2019reconstructing}. When linearly transformed fMRI signals are fed into the VAE, reliable pairwise decoding and accurate gender classification can be achieved. Compared to linear reconstruction methods, nonlinear methods, especially DNN-based methods, can greatly improve the accuracy of natural image reconstruction, especially visual details. While the performance of current decoding methods based on DNN largely depends on the size of the neural data, deep learning techniques remain one of the most promising approaches for developing visual neural decoding.

\subsection{Semi-Supervised Model}

Brain decoding suggests that image categories can be estimated through evoked neural activity. However, due to the difficulty in collecting a large amount of neural signal data for training, it is crucial to develop semi-supervised or even unsupervised decoding algorithms~\cite{akamatsu2020brain,beliy2019voxels}. \cite{akamatsu2020brain} proposed a method to accurately estimate the categories of viewed images not used for training through semi-supervised multi-view Bayesian Generative model. Their model focuses on the relationship between functional magnetic resonance imaging activity and multiple modes, namely visual features extracted from viewed images (using VGG19 to extract visual features from stimulus images) and semantic features obtained from viewed image categories (using word2vec to obtain semantic information of image categories). Moreover, in order to accurately estimate the image categories not used for training, their semi-supervised framework combines visual and semantic features obtained from other image categories besides the image categories of the training data. Specifically, VGG19 was used for the first time to extract visual features from stimulus images. In addition to training on the image categories of the data, Semi-MVBGM also utilizes visual and categorical features obtained from other image categories. Half of the MVBGM considers the unobserved fMRI activity corresponding to other visual and categorical features as missing. These missing values are initialized by interpolating random variables. Then optimize these values based on Bayesian inference method.

The estimation performance of the proposed model outperforms the state-of-the-art models in the field of brain decoding with over 95\% recognition accuracy. The results also suggest that incorporating additional image category information can improve generalization, especially when the number of training samples is small. The semi-supervised model framework is important in the field of brain decoding, where there are insufficient patterns of brain activity but sufficient visual stimuli.

\section{Conclusion }

In summary, visual information plays a central role in our ability to perceive and understand the environment. Visual information is processed through a variety of areas and regions, ultimately allowing our brains to perceive and adapt our behavior to our surroundings. Evidence points towards visual perception involving hierarchical processing and functional specialization, where visual information is transformed from simple to more abstract representations and processed by pathways specializing in different aspects of visual information. Furthermore, information is split between two main pathways: the ventral stream, responsible for perception and object recognition, and the dorsal stream, involved in guiding behavior and spatial judgment. 

The mechanisms and functions of earlier regions within the visual pathway prove simpler to understand, with latter regions often increasing in complexity of network structure, alongside increasingly sophisticated representations of visual information. Deep areas within the brain prove difficult to observe, leading to an increasing trend of computationally-driven research methods attempting to construct models mimicking aspects of the visual pathway. Deep learning models, such as CNNs, RNNs, Generative Models, GANs, and Semi-Supervised Models, have gained prominence in the field of visual object recognition and decoding of neural signals. CNNs have shown the ability to predict and decode visual stimuli by learning neural signals, while RNNs capture the temporal dynamics and feedback processing in the visual system. Generative models, like autoencoders and variational autoencoders (VAE), reconstruct visual stimuli from neural activity. GANs have been used to synthesize high-fidelity images based on brain activity measurements. Lastly, semi-supervised models aim to estimate image categories using limited neural signal data. This variety of artificial intelligence-based approaches provides promising avenues to drive future research into visual coding and our understanding of natural vision. As a return, the deep learning models provide a new approach to improve the neural decoding capability in next generation of visual neuroprotheses, and general usage of brain-machine interface.

\bibliographystyle{vancouver-modified}
\bibliography{ours}

\end{document}